\documentclass[aps,prc,showpacs,twocolumn,preprintnumbers,floatfix,
showkeys,tightenlines]{revtex4}
\usepackage{amsmath,amssymb,amsfonts}
\usepackage{graphicx}
\usepackage{color}
\allowdisplaybreaks

\newcommand{\be}{\begin{equation}}
\newcommand{\ee}{\end{equation}}
\newcommand{\bea}{\begin{eqnarray}}
\newcommand{\eea}{\end{eqnarray}}
\newcommand{\bm}{\bibitem}

\newcommand{\ep}{\epsilon}

\newcommand{\lm}{\lambda}

\newcommand{\sg}{\sigma}

\newcommand{\ze}{\zeta}

\newcommand{\cz}{{\cal Z}}

\begin{document}

\setcounter{page}{1}

\vspace*{0.5 true in}

\title{Scattering effects on nuclear thermodynamic observables}

\author{S.K. \surname{Samaddar}}
\email{santosh.samaddar@saha.ac.in}
\author{J.N. \surname{De}}
\email{jn.de@saha.ac.in}
\affiliation{Theory Division, Saha Institute of Nuclear Physics, 1/AF 
 Bidhannagar, Kolkata 700064, India}  


\begin{abstract} 
The nuclear thermodynamic observables like the temperature, freeze-out 
volume and the specific heat as obtained from isotopic ratios in 
hot disassembled nuclear matter are examined in the light of the
$S$-matrix approach to the nuclear equation of state. 
The values of the observables, as extracted without inclusion of scattering
effects are found to be modified appreciably in some cases
when scattering between the fragment species is taken care of. 
\end{abstract}


\maketitle

Understanding the nuclear equation of state (EOS) is central in clarifying
issues related to properties in compact stars, supernova dynamics or
explosive nucleosynthesis \cite{lat,she,ste,jan}. Laboratory experiments
involving energetic nuclear collisions are helpful to this end. The hot
compressed nuclear system formed in these collisions subsequently
decompresses and then becomes unstable against nuclear multifragmentation.
It is generally believed that this disassembly occurs at a density 
$\rho $ considerably lower than the saturation density of normal
nuclear matter. It is 
called the  freeze-out density where interaction among the produced
fragments are assumed to freeze out. Relation 
between the different thermodynamic state
variables like the energy $E$, temperature $T$ and the specific volume
$v$ (=1/$\rho $) of the disassembling matter  can be obtained 
from a clever manipulation of simplistic statistical
models, broadly referred to as  the nuclear statistical equilibrium (NSE)
model \cite{ran,mey}, that describe the different characteristics of the
fragments produced.

Exploiting the NSE model Albergo, Costa, Costanzo and Rubbino (ACCR)
\cite{alb} arrived at expressions for the temperature  and the
freeze-out density in terms of the experimental values of the double and
single isotope multiplicity ratios and other known quantities. The
expression for the yield $Y(A,Z)$ of a fragment of mass $A$ and charge
$Z$ in its ground state in this model is given by
\be 
Y(A,Z)=V\frac{g_A}{\lambda_A^3} e^{\left ( \mu_A+B(A,Z)\right )/T},
\ee 
where $V$ is the freeze-out volume, $T$ the temperature of the 
fragmenting system, $g_A$ the ground state degeneracy,
$\lambda_A=(\sqrt {2\pi/(AmT)}) = \lambda /\sqrt A $, $\lambda $ is
the nucleon thermal wavelength (we use natural units, $\hbar =c =1 $),
 $\mu_A$ is the chemical potential 
of the fragment and $B(A,Z)$ is its ground state binding energy.
Chemical equilibrium in the system assures that $\mu_A=N\mu_n +Z\mu_p$
where $N=A-Z$ and $\mu_n$ and $\mu_p$ are the neutron and proton
chemical potentials. The double ratio $R_2$ of the fragment yields is
then given as
\bea
R_2&&=\frac {Y(A_1^\prime,Z_1^\prime)/Y(A_1,Z_1)}{Y(A_2^\prime ,Z_2^\prime)/
Y(A_2,Z_2)}  \nonumber  \\
&&=\left (\frac{A_1^\prime A_2}{A_1A_2^\prime}\right )^{3/2}
\frac{g_{A_1^\prime}g_{A_2}}{g_{A_1}g_{A_2^\prime}}\exp (\Delta B/T).
\eea
The difference in the binding energies of the selected 
fragments  $\Delta B$ is given as
\bea
\Delta B=B(A_1^\prime,Z_1^\prime)-B(A_1,Z_1)+B(A_2,Z_2)-
B(A_2^\prime,Z_2^\prime).
\eea 
 Generally, the fragment yields $Y(A,Z)$ are selected in such a way that
$N_1^\prime =N+1, N_2^\prime = N_2+1, Z_1^\prime = Z_1, Z_2^\prime=Z_2$.
As $T$ is the only unknown quantity in Eq.(2), it can be obtained in terms
of the known quantities.

Once the temperature $T$ is thus known, the free neutron density
$\rho_n $, a measure of the volume of the fragmenting system 
is determined from the 
single isotope yield ratio $R_1=Y(A+1,Z)/Y(A,Z)$ as 
\bea
 &&\rho_n=R_1\left (\frac{A}{A+1} \right)^{3/2} \frac{g_{A}}{g_{A+1}}
\frac{2}{\lambda^3}\,\times \nonumber \\ 
&& ~~~~~\exp \{[B(A,Z)-B(A+1,Z)]/T \}.
\eea
The temperature obtained from Eq.~(2) would yield the real temperature
of the disassembling system provided the assumptions made in deriving
the NSE model are valid. This may not be so always.
One important assumption in this model is that
the fragment species produced are taken to be noninteracting
within the freeze-out volume. This
is not true except for extremely dilute systems, strong 
interaction corrections may modify the fragment multiplicities and
then the extracted temperature from the analysis as prescribed in 
Eq.~(2) differs from the real
temperature of the system. Likewise, the free neutron density extracted
through Eq.~(4) using this calculated temperature 
is not a true measure of the freeze-out volume.
A number of other assumptions like the production of the fragments
only in their ground states and the absence of any collective flow
velocity have been made in deriving Eq.~(1). The effects of these
approximations have already been studied earlier \cite{kol,shl}
and would not be considered here.

In a previous communication, it was shown that strong interaction corrections
to the NSE model \cite{mal}
can be appropriately taken up in the $S$-matrix 
approach \cite{das} to the grand partition function of the dilute
nuclear system, where in addition to all the stable mass particles,
the two-body scattering channels between them can be included 
systematically, in principle, to all orders. This is the virial 
equation of state in terms of the $S$-matrix elements. Assuming the
dilute nuclear matter to be composed only of neutrons, protons and
$\alpha $-particles, Horowitz and Schwenk \cite{hor} arrived at the 
nuclear EOS applying this methodology, it was extended later
to include $^3$H and $^3$He \cite{con}. Further extension was achieved
\cite{mal} by including all the higher mass particles and their 
excited states as well as the scattering channels formed by any
number of these species, estimated in the resonance approximation.
Inclusion of these heavier species is found to have a significant
role in the evaluation of the nuclear EOS and other quantities \cite{de}
like the nuclear symmetry energy and fragment multiplicities. 
Since the extraction of the thermodynamic
state functions like the temperature or the volume depends on the
fragment multiplicities, study of the sensitivity of these extracted
thermodynamic variables on the strong interaction effects in the 
nuclear medium is in order. The present communication aims at
this study. A somewhat related study in a different approach
has been reported \cite{shl1} recently.

The details of the $S$-matrix (SM) approach, as applied to the dilute
nuclear system is given in Ref.~\cite{mal}. For completeness, a few
relevant equations are presented below highlighting the approximations.
All the dynamical information concerning the microscopic interaction
in the system in thermodynamic equilibrium is incorporated in the
grand partition function $\cal Z$ of the system, which separates out
as \cite{das}
\be
\ln \cz =\ln {\cz}_{part} + \ln {\cz}_{scat}.
\ee
The first term corresponds to contributions from all the nuclear species
in their particle stable states behaving like an ideal quantum gas. The
second term corresponds to contributions from scattering states which
are absent in the NSE model. The first term in Eq.~(5) can further be
split into contributions from the ground and the excited states of the
bound nucleon clusters, so that 
\be 
\ln {\cz}_{part}=\ln {\cz}_{gr} +\ln {\cz}_{ex} .
\ee 
The ground state contribution is given by
\bea
&&\ln {\cz}_{gr}  =  \mp V \sum_{Z,N}g \int\!
\frac{d^3p}{(2\pi)^3}\,\times \nonumber \\ 
 && ~~~~~~~\ln \left( 1\mp \ze_{Z,N}
e^{-\beta(p^2/2Am )} \right) .
\eea
Here, the sum is over all the possible fragments including neutrons and
protons, $V$ is the volume of the system and $g$ the ground state degeneracy 
of the fragments. For $A\le 8$, the experimental values of the
degeneracies are taken, for heavier nuclei 
they are taken to be 1 for the even $A$ and 2 for odd $A$ systems. 
The fragment momentum is ${\bf p}$ and $\beta$ is the inverse temperature
$1/T$. The effective fugacity is given by $\zeta_{Z,N}=e^{\beta (\mu_{Z,N}+
B_{Z,N})}$.  $B_{Z,N}$ ($\equiv B(A,Z)$) 
is the binding energy of the fragment and $\mu_{Z,N} (\equiv \mu_A)$ is
its chemical potential.
A nucleus in a particular excited state is taken as a distinctly separate
species and can be treated in the same footing as
with the ground state. Their contributions can be obtained
by   convoluting the r.h.s. in Eq.~(7)
with their density of states in an energy interval $E_0$ to $E_s$
where $E_0$ is the first excited state taken as 2 MeV and $E_s$ is
the particle emission threshold taken as 8 MeV. 
 
The scattering term in Eq.~(5) can be formally written for nuclear matter
as \cite{mal}
\bea
\ln {\cz}_{scat}&&\!\!\!\!\!\!\!\! = V\!\sum_{Z_t,N_t}\frac{e^{\beta \mu_{Z_t,N_t}}}
{\lm^3_{A_t}}\sum_{\sg} e^{\beta B_{Z_t,N_t,\sg}} \times \nonumber \\
&&\!\!\!\!\!\!\int_0^{\infty}\!\!\!d\ep~e^{-\beta\ep} \frac{1}{2\pi i} 
Tr_{Z_t,N_t,\sg}
\left({\cal A}S^{-1}(\ep)\frac{\partial}{\partial\ep}S(\ep)\right)_c .
\eea
 Here, the double sum refers to the sum over all possible scattering
channels, formed by taking any number of particles from any of the 
stable species (excited states included) and the trace is over all
plane wave states for each of these channels. $S$ is the scattering
operator, $\cal A$ the boson symmetrization or the fermion 
antisymmetrization operator. The subscript $c$ implies only the 
connected parts of the diagrammatics of the expression in the 
parentheses and $\sigma $ denotes all other labels required to
fix a channel in this set with proton and neutron numbers $Z_t$
and $N_t$, respectively. $B_{Z_t,N_t,\sigma }$ is the sum of the
individual binding energies of all the particles in the channel
and $\ep$ is the total kinetic energy in the c.m. frame of the
scattering partners. Only two-particle scattering channels are considered
as they are expected, from binding energy considerations, to be 
more important than multi-particle channels with the same $Z_t$ and $N_t$.

For convenience, the channels are divided into light and heavy ones, 
consisting of low mass particles ($A\le 8$, say) and heavier masses,
respectively when
\be
\ln {\cz}_{scat} = \ln {\cz}_{light} + \ln {\cz}_{heavy}.
\ee
 The scattering of relatively heavier nuclei is dominated by a multitude
of resonances near the threshold; the $S$-matrix elements are then 
approximated by resonances which, like the excited states, are again
treated as ideal gas terms \cite{das1,das2}. Then, $\ln {\cz}_{heavy}$
can be written in the same form as $\ln {\cz}_{ex}$, assuming the 
level density of the resonances 
to have the same functional dependence on energy as the excited
states. We consider resonances upto excitations of 12 MeV. The damping of
the integral from the Boltzmann factor in Eq.~(8) assures contributions 
only from relatively low energies.

For the evaluation of $\ln {\cz}_{light}$, only the elastic scattering
channels for the pairs NN, Nt, N $^3$He, N $\alpha$ (N refers to the
nucleon) and $\alpha \alpha$ have been considered. These calculations
involve only experimental inputs, namely, the phase shifts and binding 
energies. Details of these
calculations can be seen in \cite{hor,de}.

Once the partition function for nuclear matter is obtained, the
pressure $P$, the number density of the $i$th species $\rho_i$,
the free energy per baryon $f$ or the entropy per baryon $s$ are
evaluated from the relations, 
\bea
P  =  T \frac{\ln {\cz}}{V}\,, 
~~~\rho_i =\ze_i\left ( \frac{\partial}{\partial \ze_i}
\frac{\ln {\cz}}{V}\right )_{V,T}, \nonumber \\
 f  =  \frac{1}{\rho} \left (\sum_i \mu_i \rho_i -P \right ),~~~~s=\frac{1}{\rho }
\left (\frac{\partial P}{\partial T}\right )_{\mu_i} . 
\eea

The total neutron and proton density are given by
\bea
\rho_n^{tot}=\sum_i N_i\rho_i \,, ~~~\rho_p^{tot}=\sum_i Z_i \rho_i,
\eea
with $\rho = \rho_n^{tot}+\rho_p^{tot}$.
The energy per baryon $e(=f+Ts)$ and the heat capacity per baryon
$C_V (=(\partial e/\partial T)_V)$ can then be obtained. 

It is evident from the above that the fragment multiplicities
obtained from the NSE model are rather approximate, the presence 
of the scattering among the different fragment species modifies
their multiplicities.
We specifically study the influence of scattering
on the extracted observables, namely, the temperature and the freeze-out
volume using double and single isotope ratios and also on the
heat capacity. For this study, we consider nuclear
matter where Coulomb is absent and only strong interaction is operative.
The sums in Eqs.~(7) and (8) run upto infinity in principle; in
practice a finite sum with a maximum mass number 
$A_{max}$=1000 is taken for 
computational facilitation. The results are found to be not very sensitive 
to further increase of $A_{max}$. The binding energies of these
nuclei are obtained using the liquid-drop type mass formula
\cite{dan} with Coulomb switched off.
To obtain the effects of scattering on the aforesaid observables,
the fragment multiplicities are calculated at a chosen temperature $T$,
baryon density $\rho $ and asymmetry
$X (=(\rho_n^{tot}-\rho_p^{tot})/\rho )$ in both the NSE model and
the SM approach. One can then infer an output temperature
and free nucleon density from the cluster composition of the system
in the two models as prescribed 
in Eqs.~(2) and (4). One easily sees that this way in the NSE model,
one gets back $T^{NSE}~=~T$, the chosen temperature. 
The ratio $T^{NSE}/T^{SM}$
and $\rho_n^{NSE}/\rho_n^{SM}= V^{SM}/V^{NSE}$ then represent
measure of the corrections due to scattering effects on the
extraction of temperature and freeze-out volume $V$ of the system
with given $\rho ,T$ and $X$.
The superscripts NSE and SM refer to quantities obtained in the NSE and
SM models, respectively. The difference in the inferred temperatures
and energies $e$
in the NSE and the SM model affects the heat capacity. Reflection of this
is shown in the difference between the specific heats at constant
volume $C_V^{SM}=(\partial e/\partial T)_V^{SM}$ and 
$C_V^{NSE}=(\partial e/\partial T)_V^{NSE}$.

The effect of scattering on  temperature  from
the double isotopic ratio is displayed in Fig.~1.
Here the ratio of $T^{NSE}$ to $T^{SM}$ is shown as a function of
the given temperature $T$ at different densities, both for symmetric ($X=0.0$)
and asymmetric ($X=0.2$) nuclear matter. The thermometer used in this
calculation is the double isotopic ratio (t/d)/($^4$He/$^3$He)
denoted as the t-He thermometer. The temperature $T^{SM}$
is always found to be lower than the NSE temperature;  
the scattering effect on the inferred temperature 
 $T^{SM}$ increases with density.
In general, at a given density, this effect has a peaked structure
which becomes more pronounced at lower densities. Isospin
asymmetry is seen to be relatively more important only at lower
temperatures, it does not affect the results qualitatively. So, 
following results are presented only for symmetric nuclear matter.

Scattering effects on the extracted temperature with the choice of the
double-ratio thermometer are presented in Fig.~2 at different densities
as a function of temperature. Three thermometers are chosen, namely,
t-He, (d/p)/($^4$He/$^3$He) (denoted d-He) and ($^4$He/$^3$He)/
($^7$Li/$^6$Li) (denoted He-Li). Choice of thermometers is seen 
to have a perceptible role on the extracted temperature, particularly
at relatively higher densities and temperatures. Unless otherwise
mentioned, the subsequent calculations refer to t-He thermometer;
among the three thermometers, it has the least scattering effect.

For very dilute warm nuclear matter, it is generally perceived that
the matter consists besides neutrons and  protons a few light species,
namely, deuterons, tritons, $^3$He and $^4$He \cite{hor,con}. 
Importance of heavier clusters surface with increasing density and
at lower temperature \cite{de}. In the density and temperature range
we explore, it is then expected that the departure of the 
temperature $T^{SM}$ from the NSE temperature $T^{NSE}$
would depend on the selection of fragment species in the model
calculation. In Fig.~3, we compare the ratio $T^{NSE}/T^{SM}$ as a
function of temperature at different densities with the choice
of light species upto $^4$He (the light species model or LSM) and
also with inclusion of the heavy species (the heavy species
model or HSM). At high temperatures, the two results tend to merge
as matter breaks up into lighter clusters. Significant difference
between the two model calculations are evident at lower temperatures
and relatively higher densities where heavier clusters are more 
abundant. These may be attributed to the fact that in the LSM,
heavier species are  absent by definition, conservation of
baryon density increases the population of the light clusters there
in comparison to those in the HSM which introduces larger scattering
effects in the LSM.

The departure of the derived value of the freeze-out volume
$V^{SM}$ from the corresponding NSE value $V^{NSE}$ due to strong interaction
effects is shown as a function of temperature in Fig.~4 through the ratio
$V^{SM}/V^{NSE}$ at  two relatively higher densities where the effects
are found to be more pronounced. Model calculations have been performed
in both HSM (upper panel) and in LSM (lower panel). The single isotope 
ratios used are (d/p) and (t/d) and the corresponding temperatures
are obtained from the thermometers (d-He) and (t-He), respectively.
The values of the $V^{SM}/V^{NSE}$ are found to be sensitive to the
choice of the single and double ratios. In LSM, the extracted
volume $V^{SM}$ differs largely from
the NSE value  at low
temperatures, but falls sharply with increase in temperature.
With inclusion of heavier species, a qualitative change in the
temperature dependence of $V^{SM}/V^{NSE}$ is observed with a marked
peak at $T \sim $ 6.0 MeV for the chosen densities. In HSM,
the estimation of the freeze-out volume could be maximally 
uncertain by a factor of $\sim $ 2 
for the cases studied.

Comparison is made between the values of the specific heat
$C_V^{SM}$ (full lines) and $C_V^{NSE}$ (dashed lines) in Fig.~5,
where they are presented at a few densities as a function of 
temperature inferred from the respective models.
The calculations at $\rho $= 0.0001, 
0.001, 0.01 and 0.02 fm$^{-3}$ are displayed in colors red, black,
magenta and cyan, respectively. The two sets of specific heats do not show
any qualitative difference. In the $S$-matrix approach, because
of scattering, the peaks in specific heat are shifted at a somewhat
lower temperature, but their magnitudes are a bit higher. 
At a particular density,
the temperature corresponding to the peak in the heat capacity reflects
the sudden onset of condensation in isochoric cooling. This is the
condensation temperature \cite{de1}. The scattering effects on the
extracted thermodynamic observables are found to be maximum here
as also can be seen from Figs.~1 and 4(a).

The ACCR method, in the context of the NSE model, gives an estimate
of the thermodynamic parameters like the temperature and the freeze-out
density of the disassembling system. They are, however, very approximate.
Various corrections are called for to improve upon these model values.
We have here explored only the effect of strong interactions among
the different species produced in the nuclear disassembly on these
thermodynamic observables. They involve model-independent entities.
Significant deviations from the ACCR values are obtained, particularly
around the condensation temperature. The extracted temperature and
the freeze-out density are always underestimated, the scattering-corrected
values can, however, be reconstructed either through iteration
or by preparing a table.

In Ref.~\cite{shl1}, modifications of the ACCR values of
the temperature and the freeze-out density are arrived at assuming
that the binding energies of the fragments in the freeze-out
configuration are dressed-up due to the presence of the medium. They
considered only light fragments and their results are qualitatively
the same as ours in the LSM. They have also incorporated effects 
due to strong interactions (they use explicit model interactions
as opposed to ours), but the connection between the two approaches 
is not immediately obvious.

\acknowledgments{The authors acknowledge the support from the Department
of Science  \& Technology, Government of India.}

\newpage
\begin{figure}[t]
\includegraphics[width=0.45\textwidth,clip=false]{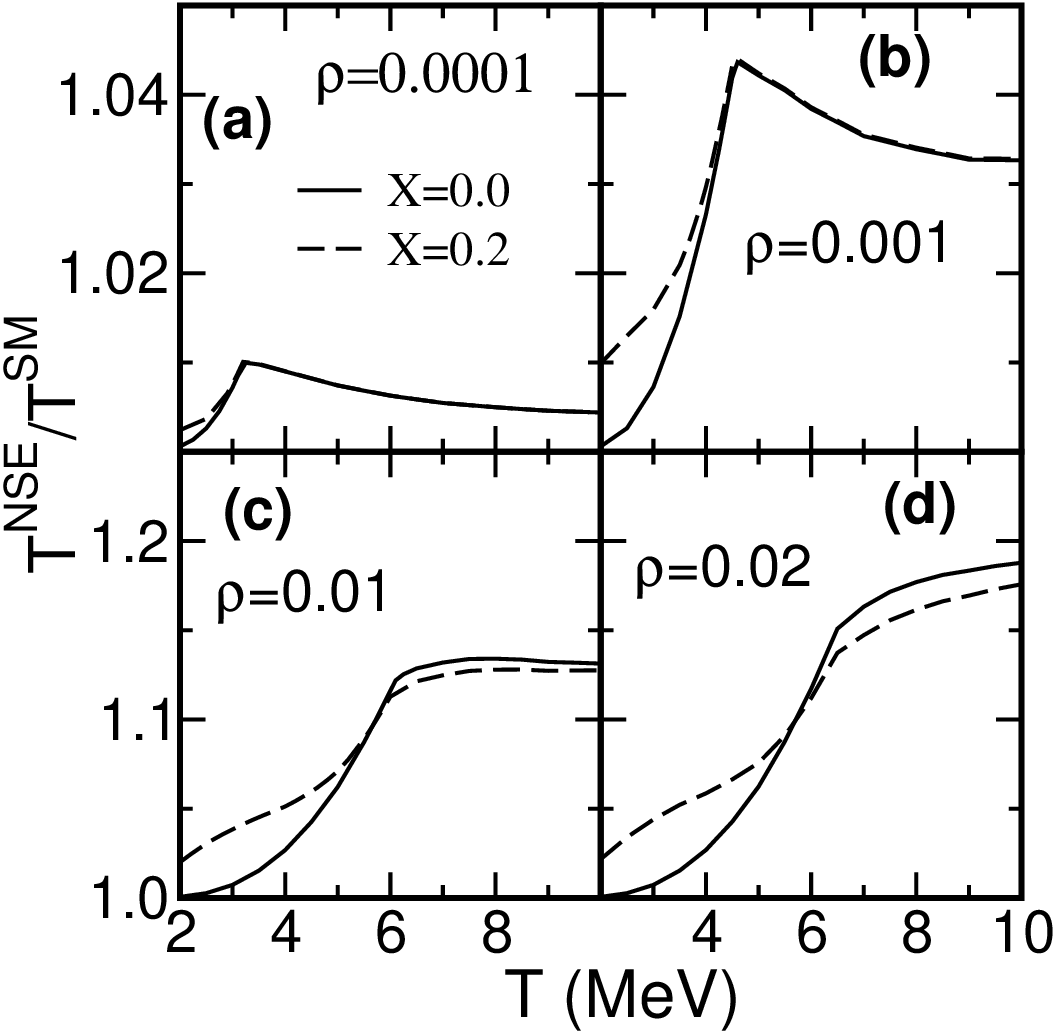}
\caption{ The ratio of the temperature
obtained in NSE to that in SM model 
using (t-He) thermometer at 
different densities $\rho $ (in fm$^{-3}$) as a function of
$T$ for symmetric ($X=0.0$, full lines) and asymmetric ($X=0.2$,
dashed lines) nuclear matter.}
\end{figure}
\begin{figure}[b]
\includegraphics[width=0.45\textwidth,clip=false]{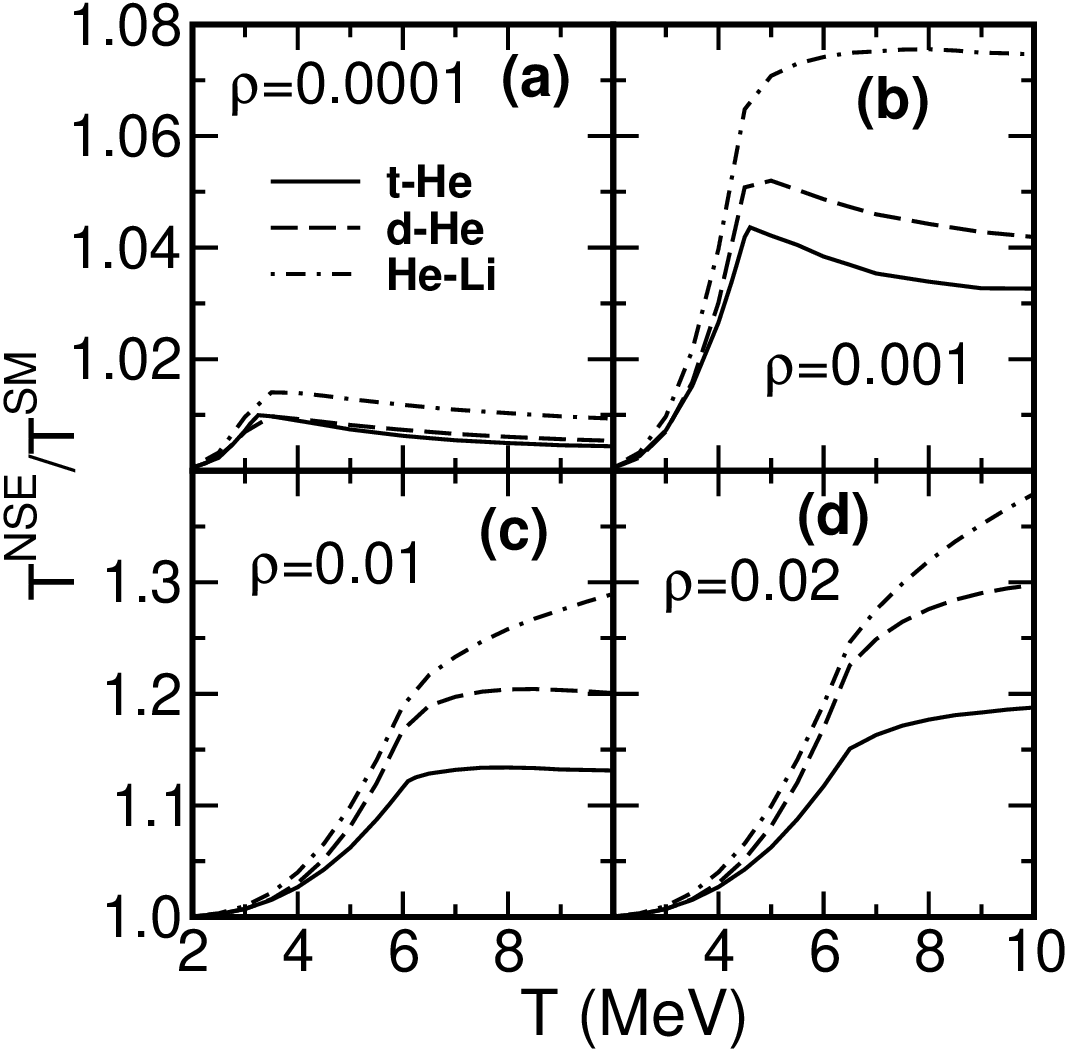}
 \caption{ Comparison of $T^{NSE}/T^{SM}$ at different densities $\rho $ 
(in fm$^{-3}$) as a function of temperature using (t-He), (d-He) and (He-Li)
thermometers shown with full, dashed and dot-dashed lines, respectively.}
\end{figure}
\begin{figure}[t]
\includegraphics[width=0.45\textwidth,clip=false]{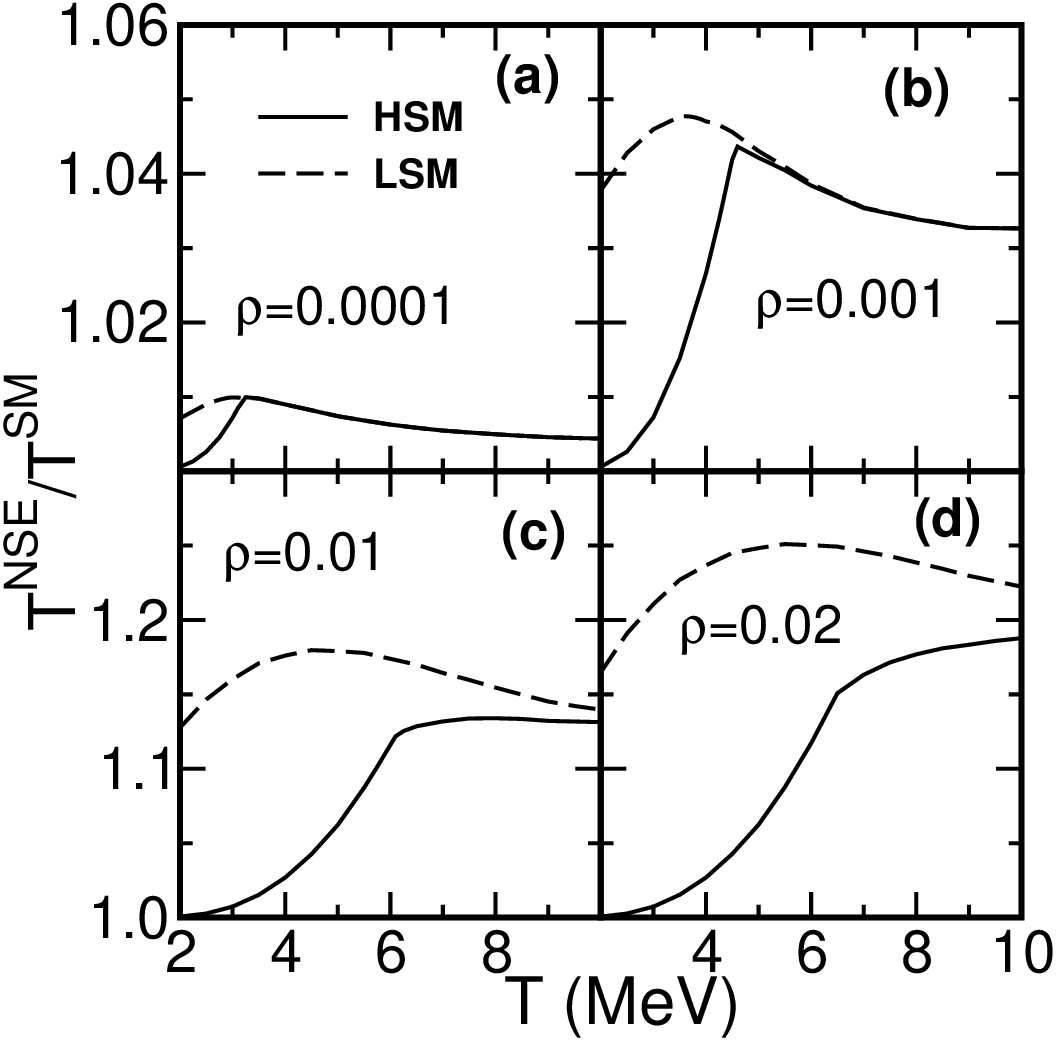}
 \caption{ The same as in Fig.~1 for symmetric nuclear matter
in HSM (full lines) and LSM (dashed lines).} 
\end{figure}
\begin{figure}[b]
\includegraphics[width=0.45\textwidth,clip=false]{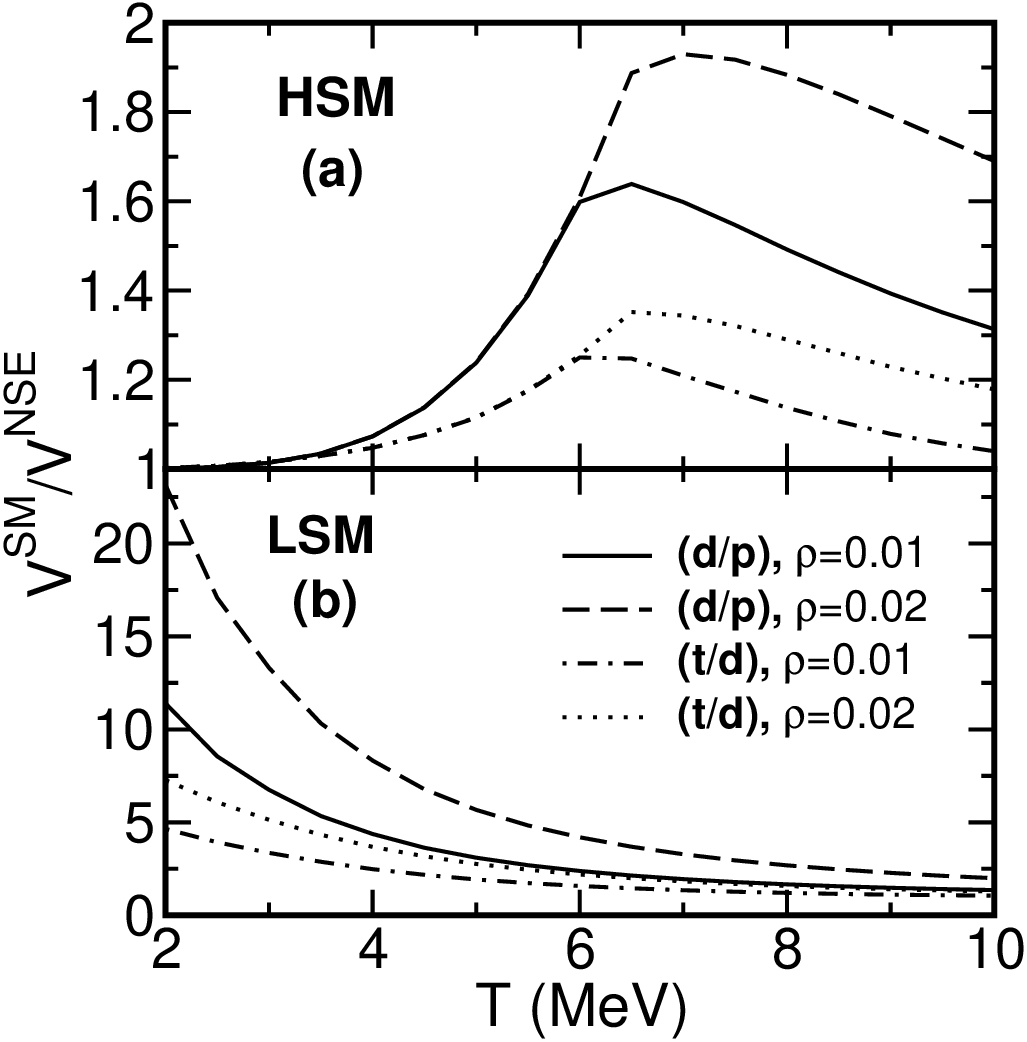}
 \caption{ The ratio of the 
freeze-out volumes obtained in SM ($V^{SM}$)
and in NSE model ($V^{NSE}$) as a function of temperature
at $\rho$=0.01 and 0.02 fm$^{-3}$ using the single ratios (d/p)
and (t/d), the corresponding thermometers are (d-He) and (t-He).}
\end{figure}
\begin{figure}[t]
\includegraphics[width=0.45\textwidth,clip=false]{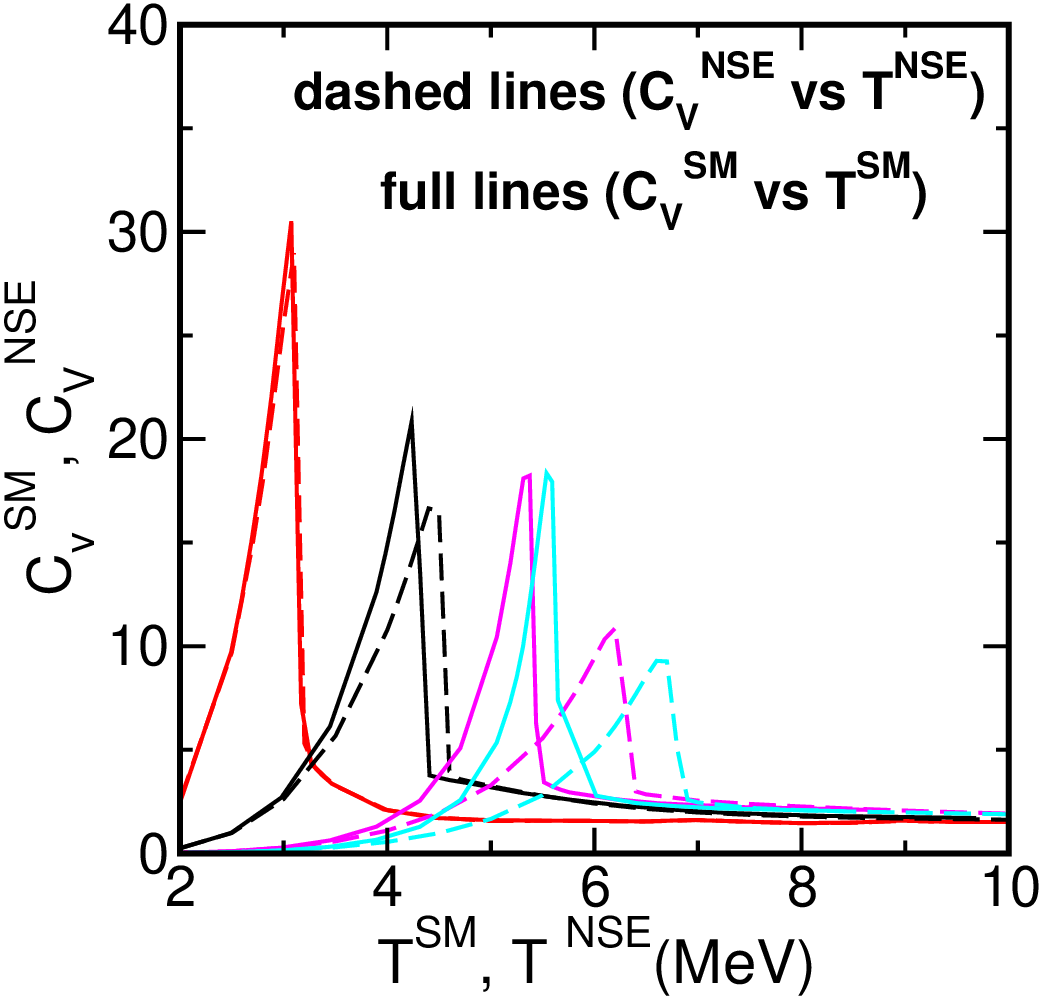}
 \caption{ (color online) 
Heat capacities at constant volume
calculated in SM ($C_V^{SM}$) and in NSE ($C_V^{NSE}$)
as a function of $T^{SM}$ and $T^{NSE}$,
respectively at different densities. The red, black, magenta and
cyan lines correspond to $\rho =$0.0001, 0.001, 0.01 and 0.02 fm$^{-3}$,
respectively.}
\end{figure}

\begin{thebibliography}{99}

\bibitem{lat} J. M. Lattimer and F. D. Swesty, Nucl. Phys.
{\bf A535}, 331 (1991).

\bibitem{she} H. Shen, H. Toki, K. Oyamatsu, and K. Sumiyoshi,
Prog. Theo. Phys. {\bf 100}, 1013 (1998).

\bibitem{ste} A. W. Steiner, M. Prakash, J. M. Lattimer, 
and P. J. Ellis,
Phys. Rep. {\bf 411}, 325 (2005).

\bm{jan} H. -Th. Janka, K. Langanke, A. Marek, G. Mart\`inez-Pinedo,
and B. M\"uller, Phys. Rep. {\bf 442}, 38 (2007).

\bm{ran} J. Randrup and S. E. Koonin, Nucl. Phys. {\bf A356},
223 (1981).

\bm{mey} B. S. Meyers, Annu. Rev. Astron. Astrophys. {\bf 32},
153 (1994).

\bm{alb} S. Albergo, S. Costa, E. Costanzo, and A. Rubbino,
Nuovo Cimento {\bf A89}, 1 (1985).

\bm{kol} A. Kolomiets {\it et al.}, Phys Rev C {\bf 54}, R472 (1996).

\bm {shl} S. Shlomo, J. N. De, and A. Kolomiets, Phys Rev C {\bf 55},
R2155 (1997).

\bm{mal} S. Mallik, J. N. De, S. K. Samaddar, and Sourav Sarkar, 
Phys. Rev. C {\bf 77}, 032201 (R) (2008).

\bm{das} R. Dashen, S-k. Ma and H.J. Bernstein, Phys. Rev, 
{\bf 187}, 345 (1969).

\bibitem{hor} C. J. Horowitz and A. Schwenk, Nucl. Phys.  {\bf A776},
55 (2006).
 

\bm{con} E. O'Connor, D. Gazit, C. J. Horowitz, A. Schwenk, and M. Barnea,
Phys. Rev. C {\bf 75}, 055803 (2007).


\bm{de} J. N. De and S. K. Samaddar, Phys. Rev. C{\bf 78},
065204 (2008).

\bm {shl1} S. Shlomo, G. R\"opke, J. B. Natowitz, L. Qin, K. Hagel, R. Wada, 
 and A. Bonasera, arXiv:0901.2036v1 [nucl-th] (2009).

\bm{das1} R. Dashen and R. Rajaraman, Phys. Rev. D {\bf 10}, 694 (1974).

\bm{das2} R. Dashen and R. Rajaraman, Phys. Rev. D {\bf 10}, 708 (1974).

\bm{dan} P. Danielewicz, Nucl. Phys. {\bf A727}, 233 (2003).

\bm{de1} J. N. De and S. K. Samaddar, Phys. Rev. C {\bf 76}, 044607 (2007).

\end{thebibliography}
\end{document}